\documentclass[twoside,a4paper,11pt]{sca}
\pdfoutput=1
\usepackage{graphicx}
\usepackage{hyperref}
\usepackage{movie15}
\usepackage{natbib}  
\begin{document}
\pagenumbering{arabic}
\pagestyle{myheadings}
\thispagestyle{empty}
{\flushright\includegraphics[width=\textwidth,bb=90 650 520 700]{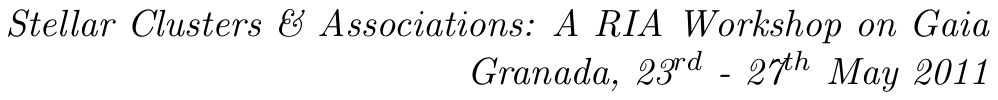}}
\vspace*{0.2cm}
\begin{flushleft}
{\bf {\LARGE
NIR view on young stellar clusters in nearby spirals
}\\
\vspace*{1cm}
P.~Grosb{\o}l$^{1}$
and 
H.~Dottori$^{2}$
}\\
\vspace*{0.5cm}
$^{1}$
European Southern Observatory, Garching, Germany\\
$^{2}$
Instituto de F\'{i}sica, Univ. Federal do Rio Grande do Sul, 
Porto Alegre, RS, Brazil\\
\end{flushleft}
\markboth{
Stellar clusters in nearby spirals}{Grosb{\o}l \& Dottori}
\thispagestyle{empty}
\vspace*{0.4cm}
\begin{minipage}[l]{0.09\textwidth}
\ 
\end{minipage}
\begin{minipage}[r]{0.9\textwidth}
\vspace{1cm}
\section*{Abstract}{\small
Observations in the near-infrared (NIR) allow a detailed study of young
stellar clusters in grand-design spiral galaxies which in visual bands often
are highly obscured by dust lanes along the arms.  Deep JHK-maps of 10 spirals
were obtained with HAWK-I/VLT. Data for NGC\,2997 are presented here to
illustrate the general results for the sample.

The (H-K)--(J-H) diagrams suggest that most stellar clusters younger than
7\,Myr are significantly attenuated by dust with visual extinctions reaching
7$^\mathrm{m}$.  A gap between younger and older cluster complexes in the
(J-K)--M$_K$ diagram indicates a rapid reduction of extinction around 7\,Myr
possibly due to expulsion of dust and gas after supernovae explosions. The
cluster luminosity function is consistent with a power law with an exponent
$\alpha \approx 2$.  Cluster luminosities of M$_k = -15^\mathrm{m}$ are
reached, corresponding to masses close to 10$^6$\,M$_\odot$, with no
indication of a cut-off.  Their azimuthal angles relative to the main spiral
arms show that the most massive clusters are formed in the arm regions while
fainter ones also are seen between the arms.  Older clusters are more
uniformly distribution with a weaker modulation relative to the arms.

\normalsize}
\end{minipage}

%
%
\section{Introduction}
Many grand-design spirals have strings of knots along their arms on
NIR maps.  Such knots have been identified as complexes of very young stellar
clusters \citep{grosbol06} which may have been triggered by a star-formation
front associated with a spiral density wave \citep{lin64, roberts69a}.  The
usage of NIR bands for the analysis of clusters provides two main advantages:
a much more complete census of young clusters, often embedded in dust lanes,
and age estimates for clusters younger than 7\,Myr.

A sample of 10 grand-design, spiral galaxies with a range of Hubble types were
selected from the study by \citet{grosbol08} and observed in the NIR to study
possible relations between spiral perturbations and star-formation.  In this
paper, we will only present results for NGC\,2997 as a representative for the
full sample.

\section{Data}
Deep maps of the galaxies were obtained in JHK-bands with the HAWK-I/VLT
instrument which provides a 7$^{\prime}$ field with 0.1$^{\prime\prime}$
pixels.  NGC\,2997 is classified as Sc(s)I.3 and has an oval distortion in its
central parts outside which a grand-design, two-armed spiral structure with a
pitch angle of 21$^\circ$ is seen.  Its distance was assumed to be 19.2\,Mpc
derived from its velocity relative to the 3K-CMB and a Hubble constant of 73
km/s/Mpc.  The reduced, stacked images had a seeing of 0.4$^{\prime\prime}$
which translates to a linear resolution of $\approx$40\,pc at the distance of
NGC\,2997.

More than 5300 sources were identified on the K-band image of NGC\,2997 using
{\it SExtractor} \citep{bertin96}.  The photometric zero points were
established through the 2MASS photometry \citep{2mass} of foreground stars in
the field.  The limiting magnitude was estimated to K$_l = 20.1^\mathrm{m}$
(i.e. M$_K$ = -11.3$^\mathrm{m}$) for 90\% completeness depending on the local
crowding.  Individual stellar clusters cannot be resolved with the resolution
of 40\,pc making it likely that many non-stellar sources are complexes of
clusters or star-forming regions.

\begin{figure}
\center
\includegraphics[scale=0.5]{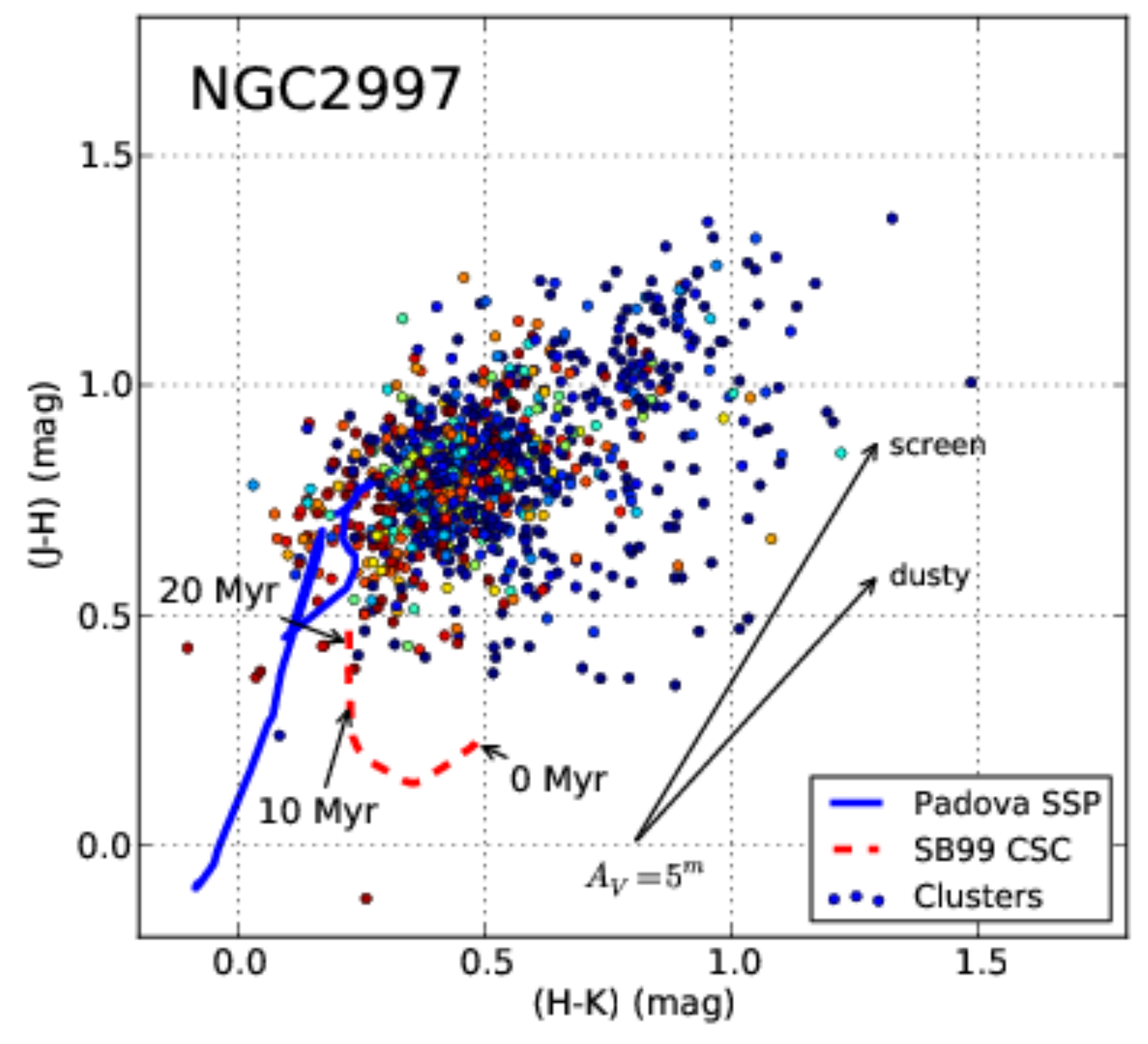}
~\includegraphics[scale=0.5]{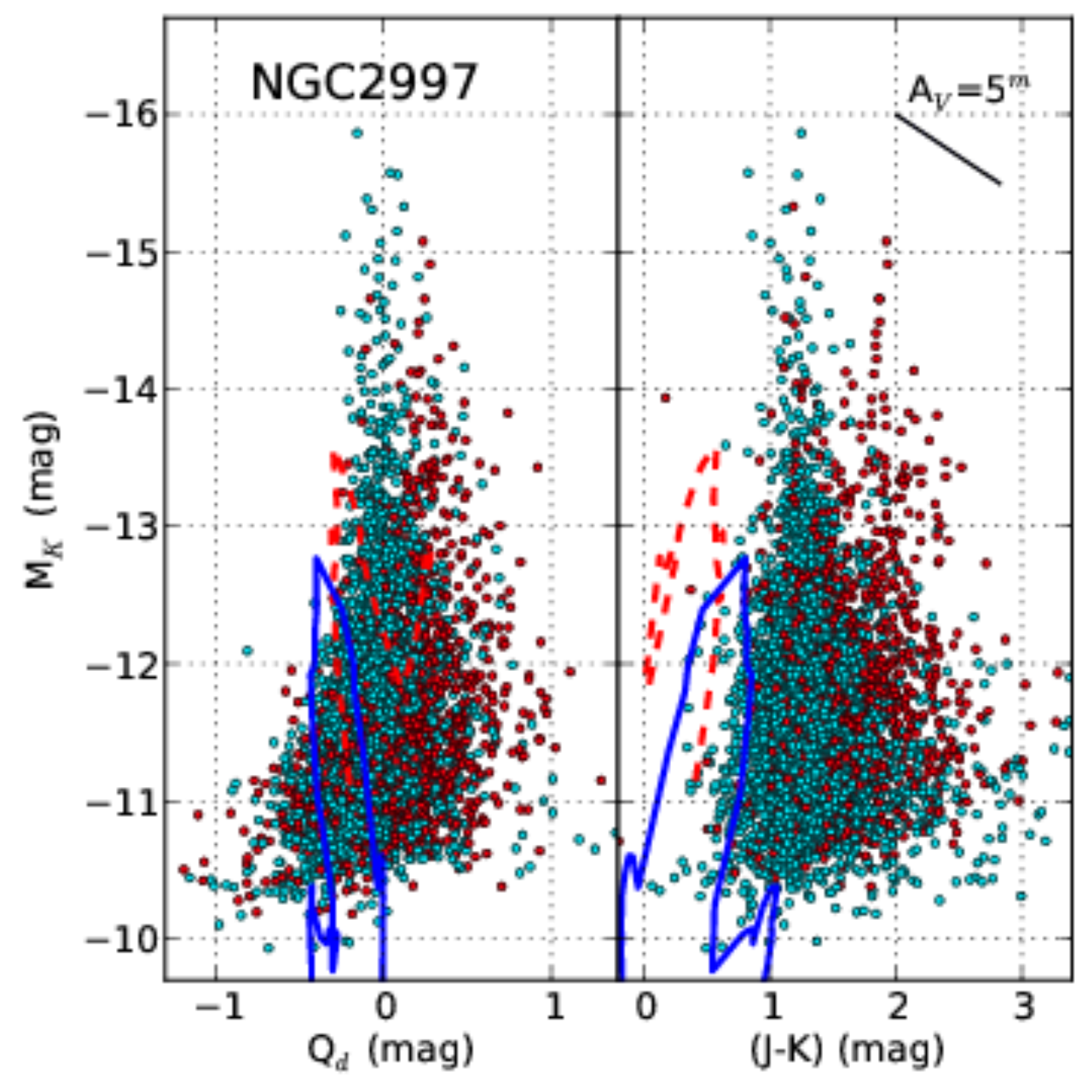} 
\caption{\label{fig1} Color-color and color-magnitude diagrams for non-stellar
sources in NGC\,2997.}
\end{figure}

\section{Color and magnitude diagrams}
The general properties of the sources can be deduced from the color-color and
color-magnitude diagrams as given in Fig.~\ref{fig1}.  The (H-K)--(J-H)
diagram for sources with photometric errors $<$0.05$^\mathrm{m}$ shows a main
group centered around (0.4, 0.8) being consistent with colors of somewhat
reddened, stellar clusters with ages $>$50\,Myr.  Evolutionary tracks for a
single-burst stellar population (SSP) using the Padova isochrones
\citep{marigo08} and a continuous star-forming cluster (CSC) from Starburst99
\citep[ hereafter SB99]{leitherer99} are indicated on the figure.  The
reddening vectors for a screen model \citep{indebetouw05} and a dusty
environment \citep{witt92, israel98} are also plotted for a visual extinction
of 5$^\mathrm{m}$.  A second, smaller group is located close to (0.8, 1.1)
which is likely to contain highly reddened, dusty young clusters.  Finally,
there is a scatter of sources to redder (H-K) values which may be caused by
emission from hot dust.  Whereas many older cluster has small extinction, most
young complexes have several magnitudes of visual extinction.

A reddening correction color index Q$_d$ = (H-K) - 0.84$\times$(J-H) can be
constructed using the dusty models of \citet{witt92}.  The color-magnitude
diagrams for the absolute magnitude M$_K$ is shown on the right in
Fig.~\ref{fig1} as functions of Q$_d$ and (J-K) indexes. Tracks of
evolutionary SSP models are indicated on the figures together with a dusty
reddening vector.  The Q$_d$-M$_K$ diagram is consistent with the tracks with
some (H-K) excess due to hot dust and absolute magnitudes up to M$_K \approx
-15^\mathrm{m}$ corresponding to masses of the order of 10$^6$\,M$_\odot$ with
a Salpeter IMF and an upper mass limit of 100\,M$_\odot$.  The (J-K) plot
shows a double branch structure where the redder one with (J-K) =
1.9$^\mathrm{m}$ consists of the young, dusty clusters in the color-color
diagram while the older clusters are located around 1.2$^\mathrm{m}$.  From
the evolutionary tracks, it is expected that intrinsic (J-K) colors of
clusters become redder with age.  The gap between the two branches suggests a
rapid reduction of extinction in the clusters which may be caused by expulsion
of dust and gas due to supernovae explosions.

\begin{figure}
\center
\includegraphics[scale=0.5]{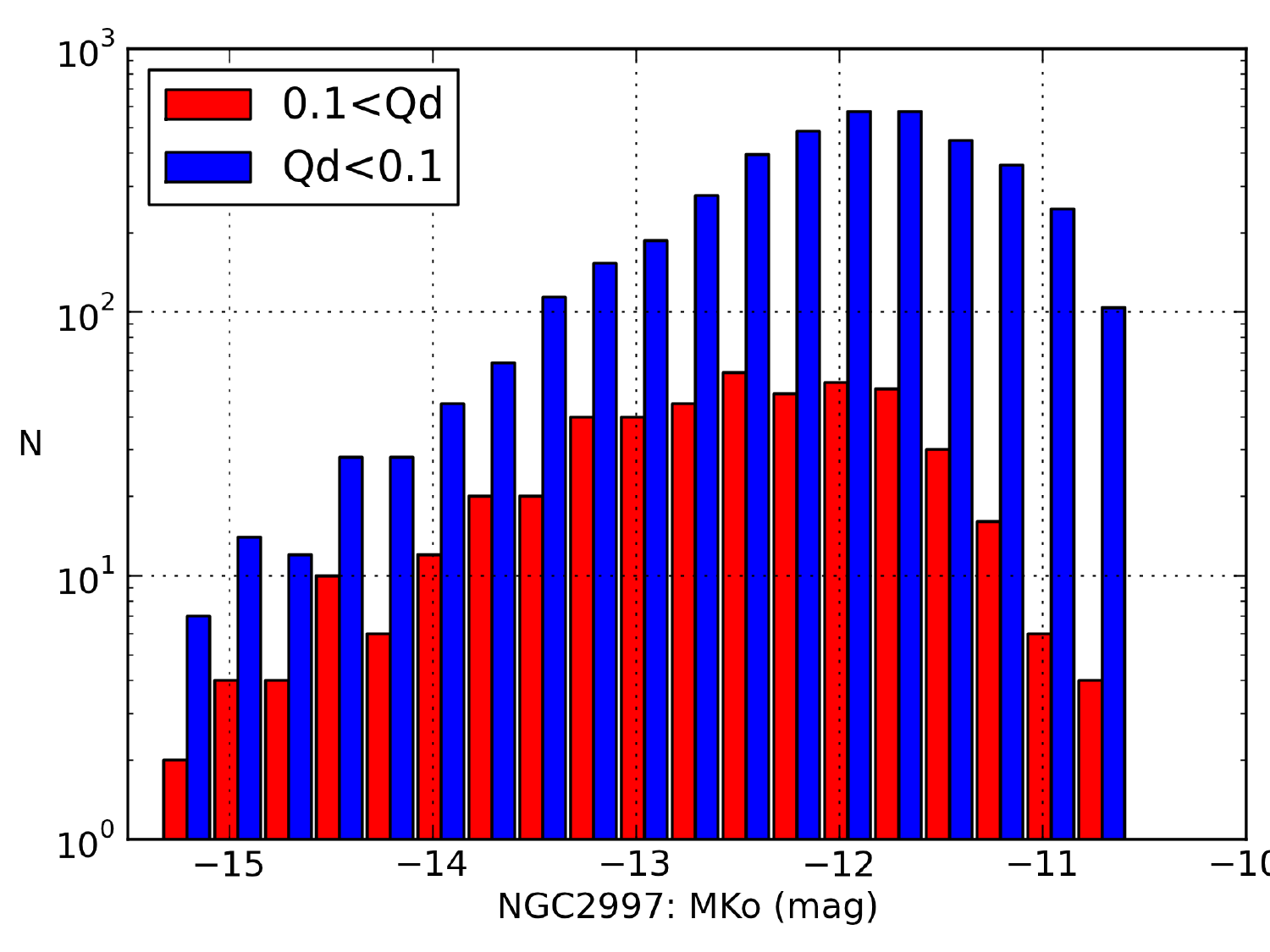}~
\includegraphics[scale=0.5]{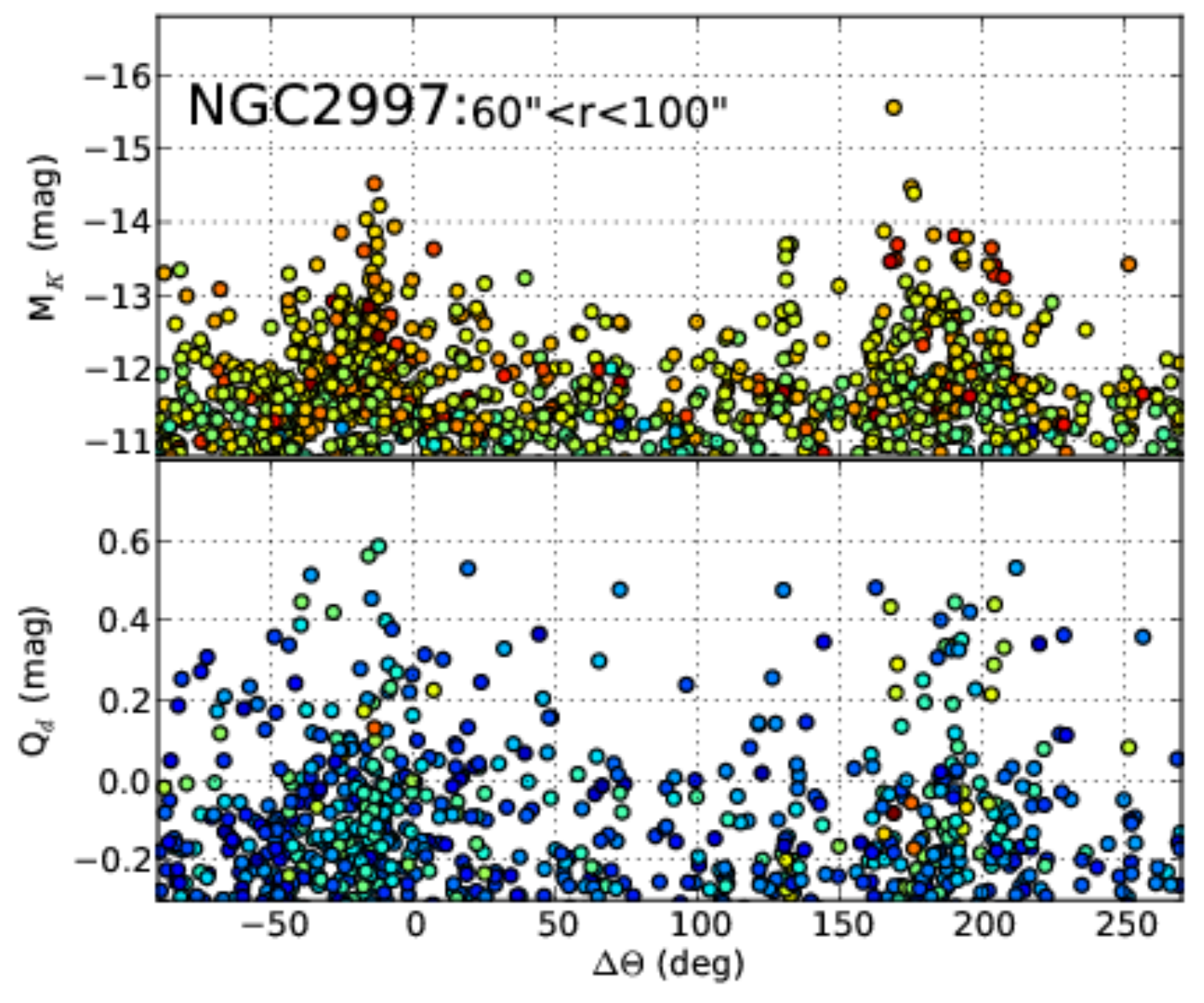} 
\caption{\label{fig2} Luminosity function (left) and azimuthal distribution
  relative to the main two-armed spiral in the radial region
  $60^{\prime\prime} - 100^{\prime\prime}$ (right) of cluster complexes in
  NGC\,2997. }
\end{figure}

\section{Luminosity and spatial distribution}
The attenuation by dust can be estimated from the (H-K)--(J-H) diagram using
the SB99 model track for the intrinsic colors. Applying this correction, the
distribution of the 'reddening free' absolute magnitude M$_{Ko}$ is shown in
Fig.~\ref{fig2} where the sources are separated in young complexes with
0.1$^\mathrm{m} <$Q$_d$ (i.e. age $<$7\,Myr) and older ones. Incompleteness
starts to be important at -12$^\mathrm{m}<$M$_{Ko}$ for the older population
while the young one is affected already around -13$^\mathrm{m}$ due to their
higher average extinction.  The high luminosity part of both samples follows a
power law with almost the same exponent $\alpha$ = 2.1 and no significant
indication of a cut-off at brightest magnitudes.  The ratio between the two
populations varies significantly from galaxy to galaxy in the sample and
reflects the current star-formation rate relative to the average.

The main symmetric part of the spiral pattern in NGC\,2997 starts around
60$^{\prime\prime}$, just outside the oval distortion, and breaks up close to
100$^{\prime\prime}$.  The phase of the arms in this region was determined
from the m=2 Fourier component of the azimuthal K-band intensity variation in
1$^{\prime\prime}$ radial bins.  Absolute magnitude M$_K$ and color index
Q$_d$ of complexes are plotted as function of their azimuthal angle
$\Delta\Theta$ relative to the main arms in the right panel of
Fig.~\ref{fig2}.  The M$_K$ values are clearly peaked in the arm regions
(i.e. around 0$^\circ$ and 180$^\circ$) with very few source brighter than
-13$^\mathrm{m}$ between the arms.  A similar picture is seen for
$\Delta\Theta$-Q$_d$ diagram where young complexes with $0.1^\mathrm{m}<$Q$_d$
are more frequent in the arms although some fainter clusters are formed in
between the arms.  Older, fainter complexes are more uniformly distributed
with only a moderate increase in density in the arm regions.

\section{Conclusions}
In a sample of 10 nearby grand-design, spiral galaxies, NGC\,2997 was selected
to illustrate the general properties of their stellar cluster populations as
observed in the NIR.  The cluster complexes form two distinct groups in the
(H-K)--(J-H) diagram where the larger is consistent with an older stellar
population with low extinction while the smaller mainly contains of young
complexes with ages $<$7\,Myr and visual extinctions in the range
2-7$^\mathrm{m}$.  No young cluster with A$_V<1^\mathrm{m}$ was found while
some are scattered to redder (H-K) colors possibly by emission from hot dust.
A gap between older and younger complexes in (J-K)-M$_K$ diagram suggests a
rapid expulsion of dust at an age around 7\,Myr which could be triggered by
supernovae explosions.

The upper part of the cluster luminosity function is well fitted by a power
law with an index $\alpha \approx 2$ with no indication of a cut-off at the
bright end.  The most luminous complexes with M$_{Ko}$ around -15$^\mathrm{m}$
may have masses close to 10$^6$\,M$_\odot$ assuming a Salpeter IMF.  The
distribution of sources relative to the spiral arms indicates that the most
massive clusters predominantly are formed in the arm regions while fainter and
older cluster show a more uniform azimuthal distribution with a weaker
modulation relative to the arms.

%
\small  
%
%

%
%
%
%

\bibliographystyle{aa}
\bibliography{AstronRef}

\begin{thebibliography}{11}
\expandafter\ifx\csname natexlab\endcsname\relax\def\natexlab#1{#1}\fi

\bibitem[{Bertin \& Arnouts(1996)}]{bertin96}
Bertin, E. \& Arnouts, S. 1996, A\&AS, 117, 393

\bibitem[{Grosb{\o}l \& Dottori(2008)}]{grosbol08}
Grosb{\o}l, P. \& Dottori, H. 2008, A\&A, 490, 87

\bibitem[{Grosb{\o}l {et~al.}(2006)Grosb{\o}l, Dottori, \& Gredel}]{grosbol06}
Grosb{\o}l, P., Dottori, H., \& Gredel, R. 2006, A\&A, 453, L25

\bibitem[{Indebetouw {et~al.}(2005)Indebetouw, Mathis, Babler, Meade, Watson,
  Whitney, {et~al.}}]{indebetouw05}
Indebetouw, R., Mathis, J.~S., Babler, B.~L., {et~al.} 2005, ApJ, 619, 931

\bibitem[{Israel {et~al.}(1998)Israel, van~der Werf, Hawarden, \&
  Aspin}]{israel98}
Israel, F.~P., van~der Werf, P.~P., Hawarden, T.~G., \& Aspin, C. 1998, A\&A,
  336, 433

\bibitem[{Leitherer {et~al.}(1999)Leitherer, Schaerer, Goldader, {Gonz{\'a}lez
  Delgado}, Robert, {et~al.}}]{leitherer99}
Leitherer, C., Schaerer, D., Goldader, J.~D., {et~al.} 1999, ApJS, 123, 3

\bibitem[{Lin \& Shu(1964)}]{lin64}
Lin, C.~C. \& Shu, F.~H. 1964, ApJ, 140, 646

\bibitem[{Marigo {et~al.}(2008)Marigo, Girardi, Bressan, Groenewegen, Silva, \&
  Granato}]{marigo08}
Marigo, P., Girardi, L., Bressan, A., {et~al.} 2008, A\&A, 482, 883

\bibitem[{Roberts(1969)}]{roberts69a}
Roberts, W.~W. 1969, ApJ, 158, 123

\bibitem[{Skrutskie {et~al.}(2006)Skrutskie, Cutri, Stiening, Weinberg,
  Schneider, Carpenter, {et~al.}}]{2mass}
Skrutskie, M.~F., Cutri, R.~M., Stiening, R., {et~al.} 2006, AJ, 131, 1163

\bibitem[{Witt {et~al.}(1992)Witt, Thronson, \& {Capuano, Jr.}}]{witt92}
Witt, A.~N., Thronson, H.~A., \& {Capuano, Jr.}, J.~M. 1992, ApJ, 393, 611

\end{thebibliography}

\end{document}